# Comment: Microarrays, Empirical Bayes and the Two-Group Model

**T. Tony Cai**

Professor Efron is to be congratulated for his innovative and valuable contributions to large-scale multiple testing. He has given us a very interesting article with much material for thought and exploration. The two-group mixture model (2.1) provides a convenient and effective framework for multiple testing. The empirical Bayes approach leads naturally to the local false discovery rate (Lfdr) and gives the Lfdr a useful Bayesian interpretation. This and other recent papers of Efron raised several important issues in multiple testing such as theoretical null versus empirical null and the effects of correlation. Much research is needed to better understand these issues.

Virtually all FDR controlling procedures in the literature are based on thresholding the ranked $p$-values. The difference among these methods is in the choice of the threshold. In multiple testing, typically one first uses a $p$-value based method such as the Benjamini–Hochberg procedure for global FDR control and then uses the Lfdr as a measure of significance for individual nonnull cases. See, for example, Efron (2004, 2005). In what follows I will first discuss the drawbacks of using $p$-value in large-scale multiple testing and demonstrate the fundamental role played by the Lfdr. I then discuss estimation of the null distribution and the proportion of the nonnulls. I will end with some comments about dealing with the dependency.

In the discussion I shall use the notation given in Table 1 to summarize the outcomes of a multiple testing procedure.

*T. Tony Cai is Dorothy Silberberg Professor of Statistics, Department of Statistics, The Wharton School, University of Pennsylvania, Philadelphia, Pennsylvania 19104, USA (e-mail: tcai@wharton.upenn.edu).*

With the notation given in the table, the false discovery rate (FDR) is then defined as FDR = $\mathrm{E}(N_{10}/R|R>0)\mathrm{Pr}(R>0)$.

## 1. THE USE OF p-VALUES: VALIDITY VERSUS EFFICIENCY

In the classical theory of hypothesis testing the $p$-value is a fundamental quantity. For example, the decision of a test can be made by comparing the $p$-value with the prespecified significance level $\alpha$. In the more recent large-scale multiple testing literature, $p$-value continues to play a central role. As mentioned earlier, nearly all FDR controlling procedures separate the nonnull hypotheses from the nulls by thresholding the ordered $p$-values.

A dual quantity to the false discovery rate is the false nondiscovery rate $\mathrm{FNR} = \mathrm{E}(N_{01}/S|S>0) \times \mathrm{Pr}(S>0)$. In a decision-theoretical framework, a natural goal in multiple testing is to find, among all tests which control the FDR at a given level, the one which has the smallest FNR. We shall call an FDR procedure *valid* if it controls the FDR at a prespecified level $\alpha$, and *efficient* if it has the smallest FNR among all FDR procedures at level $\alpha$. The literature on FDR controlling procedures so far has focused virtually exclusively on the validity; the efficiency issue has been mostly untouched.

In a recent article, Sun and Cai (2007) considered the multiple testing problem from a compound decision point of view. It is demonstrated that $p$-value is in fact not a fundamental quantity in large-scale multiple testing; the local false discovery rate (Lfdr) is. Thresholding the ordered $p$-values does not in general lead to efficient multiple testing procedures. The reason for the inefficiency of the $p$-value methods can be traced back to Copas (1974) where a weighted classification problem was considered. Copas (1974) showed that if a symmetric classification rule for dichotomies is admissible, then it must be ordered by the likelihood ratios, which is equivalent to being ordered by the Lfdr. Sun and Cai (2007) showed that, under mild conditions, the multiple testing problem is in fact equivalent to the weighted classification problem. I will discuss below some of the findings in Sun and Cai (2007) and draw connections to the present paper by Professor Efron.

TABLE 1

|         | Claimed nonsignificant | Claimed significant | Total |
|---------|------------------------|---------------------|-------|
| Null    | $N_{00}$               | $N_{10}$            | $m_0$ |
| Nonnull | $N_{01}$               | $N_{11}$            | $m_1$ |
| Total   | $S$                    | $R$                 | $m$   |

The local false discovery rate, defined in (2.7), was first introduced in Efron et al. (2001) as the a posteriori probability of a gene being in the null group given the $z$-score $z$. The results in Sun and Cai (2007) show that the Lfdr is a fundamental quantity which can be used directly for optimal FDR control. By using the Lfdr directly for testing, the goals of global error control and individual case interpretation are naturally unified.

For convenience, in the following we shall work with the marginal false discovery rate mFDR = $\mathrm{E}(N_{10})/\mathrm{E}(R)$ and the marginal false nondiscovery rate mFNR = $\mathrm{E}(N_{01})/\mathrm{E}(S)$. The mFDR is asymptotically equivalent to the usual FDR under weak conditions, mFDR = FDR + $O(m^{-1/2})$, where $m$ is the number of hypotheses. See Genovese and Wasserman (2002).

It is illustrative to first look at an example in the so-called oracle setting, where in the two-group mixture model (2.6) the proportion $p_0$, the density $f_0$ of the null distribution and the density $f$ of the marginal distribution are assumed to be known. In this case, both the optimal threshold for the $p$-values and the optimal threshold for the Lfdr values can be calculated for any given mFDR level. We shall call a testing procedure with the optimal cutoff the *oracle procedure*. Suppose the $z$-values $z_1, \ldots, z_m$ come from a normal mixture distribution with

$$(1) \quad f(z) = p_0 \phi(z) + p_1 \phi(z - \mu_1) + p_2 \phi(z - \mu_2),$$

where $p_0 = 0.8$, $p_1 + p_2 = 0.2$. That is, in the two-group model (2.6), the null distribution is $N(0,1)$, the distribution of the nonnulls is a two-component normal mixture, and the total proportion of the nonnulls is 0.2. Figure 1 compares the performance of the $p$-value and Lfdr oracle procedures (see Sun and Cai, 2007).

In Figure 1, panel (a) plots the mFNR of the two oracle procedures as a function of $p_1$ in (1) where the mFDR level is set at 0.10, and the means under the alternative are $\mu_1 = -3$ and $\mu_2 = 3$. Panel (b) plots the mFNR as a function of $p_1$ in the same setting except that the alternative means are $\mu_1 = -3$ and $\mu_2 = 6$. In panel (c) we choose mFDR = 0.10, $p_1 = 0.18$, $p_2 = 0.02$, $\mu_1 = -3$ and plot the mFNR as a function of $\mu_2$. Panel (d) plots the mFNR as a function of the mFDR level while holding $\mu_1 = -3$, $\mu_2 = 1$, $p_1 = 0.02$, $p_2 = 0.18$ fixed.

It is clear from the plots that the $p$-value oracle procedure is dominated by the Lfdr oracle procedure. At the same mFDR level, the mFNR of the Lfdr oracle procedure is uniformly smaller than the mFNR of the $p$-value oracle procedure. The largest difference occurs when $|\mu_1| < \mu_2$ and $p_1 > p_2$, where the alternative distribution is highly asymmetric about the null. When $|\mu_1| = |\mu_2|$, the mFNR remains a constant for the $p$-value oracle procedure, while the mFNR for the Lfdr oracle procedure can be noticeably smaller when $p_1$ and $p_2$ are significantly different, in which case the nonnull distribution has a high degree of asymmetry. The Lfdr oracle procedure utilizes the distributional information of the nonnulls, but the $p$-value oracle procedure does not.

The Lfdr oracle procedure ranks the relative importance of the test statistics according to their likelihood ratios. An interesting consequence of using the Lfdr statistic in multiple testing is that an observation located farther from the null (i.e., a larger absolute $z$-value or equivalently a smaller $p$-value) may have a lower significance level. It is therefore possible that the test accepts a more "extreme" observation while rejecting a less extreme observation, which implies that the rejection region is asymmetric. This is not possible for a testing procedure based on the individual $p$-values, whose rejection region is always symmetric about the null. This can be seen from Figure 2. The left panel compares the mFNR of the $p$-value oracle procedure and Lfdr oracle procedure and the right panel compares the rejection region in the case of $p_1 = 0.15$. In this case the Lfdr procedure rejects a $z$-value of $-2$ (Lfdr = 0.227, $p$-value = 0.046) but not a $z$-value of 3 (Lfdr = 0.543, $p$-value = 0.003). More numerical results are given in Sun and Cai (2007). The results show that the Lfdr oracle procedure dominates the $p$-value procedure in all configurations of the nonnull hypotheses.

The difference between the two procedures can be even more striking when the alternative distribution $f_1$ is highly concentrated. In this setting, it is possible that the extreme $p$-values near both 0 and 1 actually all come from the null distribution instead of the nonnull distribution! In such a case, thresholding the $p$-values fails completely as a method for separating the nonnull hypotheses from the nulls. In contrast, the Lfdr can still be effective in distinguishing between the null and nonnull cases.

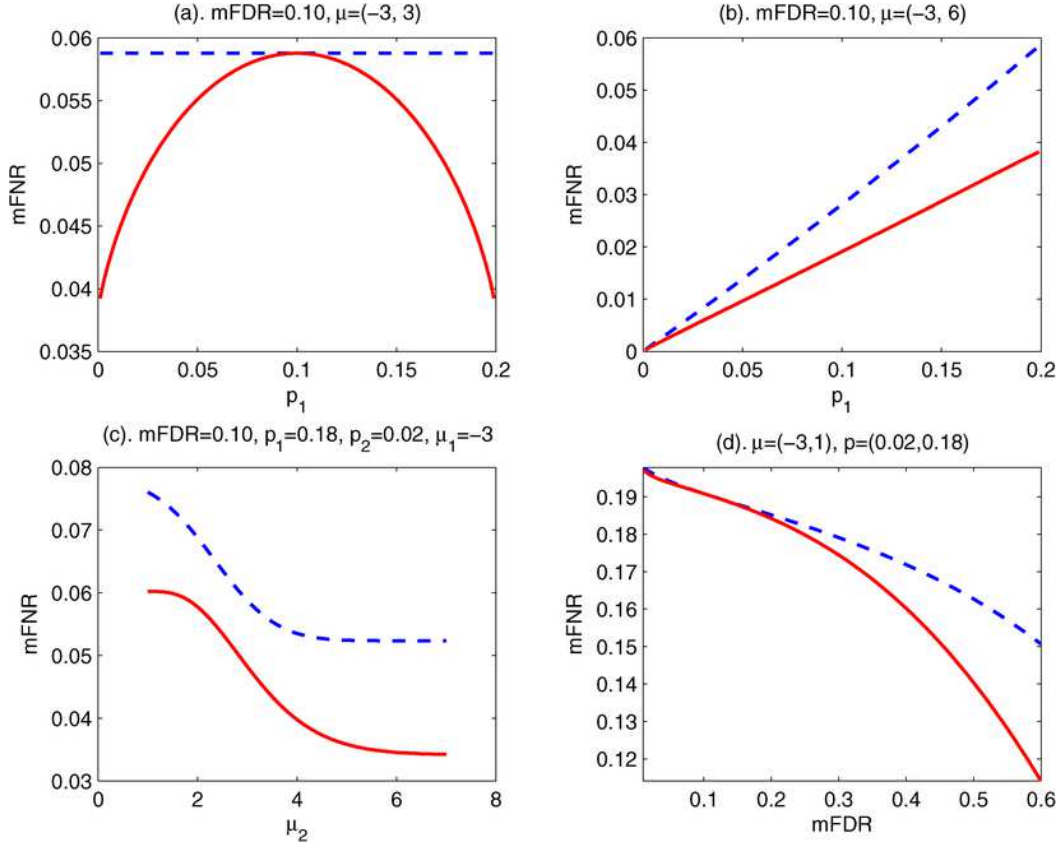

Fig. 1. *The comparison of the p-value (dashed line) and z-value (solid line) oracle rules.*

In real applications, the proportion $p_0$ and the density of the marginal distribution $f$ are unknown. With a large number of observed $z$-values, both $p_0$ and $f$ can be estimated well from the data. In this regard, the large-scale nature of the problem is a blessing. The null distribution is more subtle. If all the mathematical assumptions are satisfied, the theoretical null distribution is true and thus can be used to compute the Lfdr values. Otherwise, as argued convincingly by Efron in Section 5 of the present paper, the empirical null distribution should be used and it can be estimated from the data. Among the three quantities, $p_0$, $f_0$ and $f$, the marginal density $f$ is relatively easier to estimate than $p_0$ and $f_0$. Optimal estimation of these quantities is a challenging problem. We shall discuss the estimation issue in the next section. Let us assume for the moment that we already have consistent estimators $\hat{p}_0$, $\hat{f}_0$ and $\hat{f}$. Such consistent estimators are provided, for example, in Jin and Cai (2007). Define the estimated Lfdr by $\widehat{\mathrm{Lfdr}}(z_i) = [\hat{p}_0 \hat{f}_0(z_i)/\hat{f}(z_i)] \wedge 1$. Sun and Cai (2007) introduced the following adaptive step-up procedure:

$$\text{Let } k = \max\left\{i : \frac{1}{i}\sum_{j=1}^{i} \widehat{\mathrm{Lfdr}}_{(j)} \leq \alpha\right\},$$

(2)

$$\text{then reject all } H_{(i)},\ i=1,\ldots,k.$$

It was shown that the data-driven procedure (2) controls the mFDR at level $\alpha$ asymptotically and the mFNR level of the adaptive procedure (2) is asymptotically equal to the mFNR level achieved by the Lfdr oracle procedure. In this sense, the adaptive procedure (2) is asymptotically efficient. Numerical studies in Sun and Cai (2007) show that this adaptive procedure outperforms the step-up procedure (Benjamini and Hochberg, 1995) and the adaptive $p$-value based procedure (Benjamini and Hochberg, 2000; Genovese and Wasserman, 2004). The numerical results are consistent with the theoretical argu-ments. These results together show that the Lfdr, not the $p$-value, is a fundamental quantity for large-scale multiple testing.

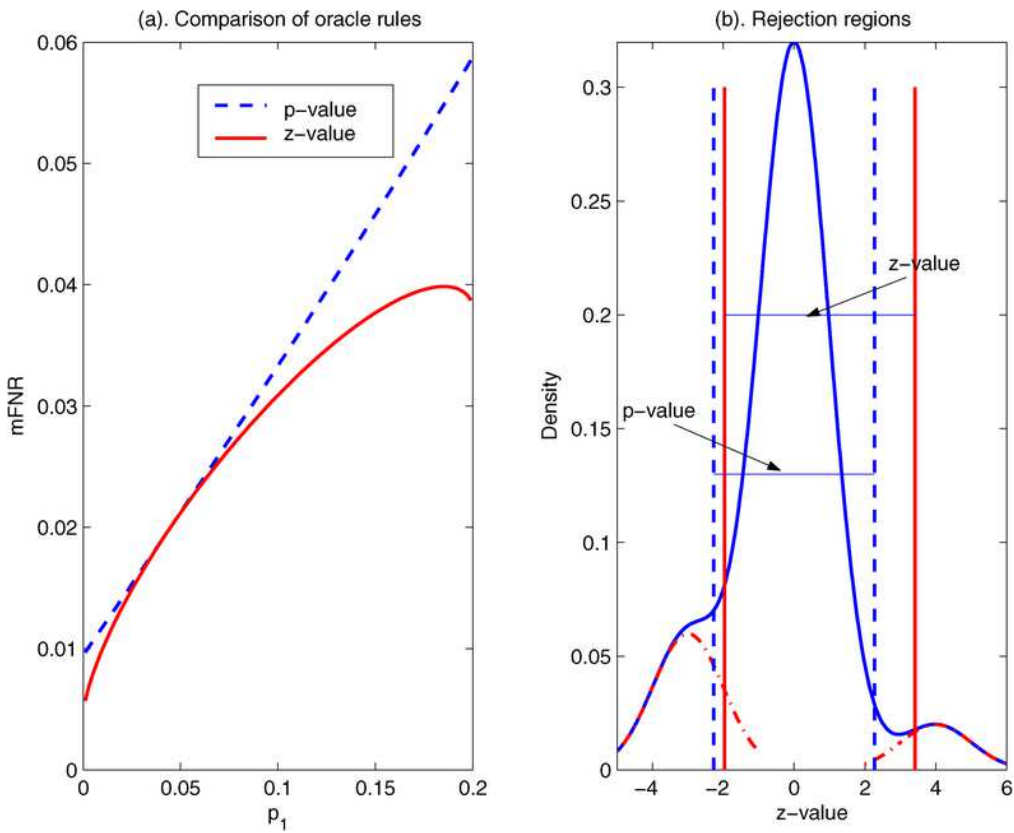

FIG. 2. *Symmetric rejection region versus asymmetric rejection region. In the mixture model (1), $\mu_1 = -3$ and $\mu_2 = 4$. Both procedures control the mFDR at 0.10.*

It is clear that the performance of the adaptive testing procedure (2) depends to a certain extent on the estimation accuracy of the estimators $\hat{p}_0$, $\hat{f}_0$ and $\hat{f}$. This leads to the estimation issue, which will be discussed next.

## 2. ESTIMATING THE NULL DISTRIBUTION AND THE PROPORTION OF THE NONNULLS

As demonstrated convincingly in this and other recent papers of Efron, the true null distribution of the test statistic can be quite different from the theoretical null and two seemingly close choices of the null distribution can lead to substantially different testing results. This demonstrates that the problem of estimating the null density $f_0$ is important to simultaneous multiple testing. In addition to the null density $f_0$, the proportion of the nonnulls is another important quantity.

Conventional methods for estimating the null parameters are based on either moments or extreme observations. In the present paper, two methods, analytical and geometric, for estimating the null density are discussed. In addition, Efron (2004) suggested an approach which uses the center and half width of the central peak of the histogram for estimating the parameters of the null distribution. These methods are convenient to use. However, the properties of these estimators are still mostly unknown. For example, the analytical method appears to be quite sensitive to the choice of the interval $[a, b]$. It is interesting to understand how the choice of $[a, b]$ affects the resulting estimator $\hat{f}_0$, and more importantly the outcomes of the subsequent testing procedures.

The three null density estimation methods mentioned above rely heavily on the sparsity assumption which means that the proportion of nonnulls is small and most of the $z$-values near zero come from the nulls. In the nonsparse case these methods of estimating the null densities do not perform well and it is not hard to show that the estimators are generally inconsistent.

Jin and Cai (2007) introduced an alternative frequency domain approach for estimating the null parameters by using the empirical characteristic function and Fourier analysis. The approach demonstrates that the information about the null is well preserved in the high-frequency Fourier coefficients, where the distortion of the nonnull effects is asymptotically negligible. The approach integrates the strength of several factors, including sparsity and heteroscedasticity, and provides good estimates of the null in a much broader range of situations than existing approaches do. The resulting estimators are shown to be uniformly consistent over a wide class of parameters and outperform existing methods in simulations. The approach of Jin and Cai (2007) also yields a uniformly consistent estimator for the proportion of nonnull effects. In a two-component normal mixture setting, Cai, Jin and Low (2007) proposed an estimator of the proportion and developed a minimax theory for the estimation problem.

Much research is still needed in this area. In particular, it is of significant interest to understand how well the null density can be estimated and how the performance of the estimators affects the performance of the subsequent multiple testing procedures.

## 3. MODELING THE DEPENDENCY

This paper also raised the important issue of the effects of correlation on outcomes of the testing procedures. Observations arising from large-scale multiple comparison problems are often dependent. For example, different genes may cluster into groups along biological pathways and exhibit high correlation in microarray experiments. It is noted in this paper that correlation can considerably widen or narrow the null distribution of the $z$-values, and so must be accounted for in deciding which hypotheses should be reported as nonnull. In fact, the notion of null distribution itself becomes unclear in the dependent case.

The focus of previous research on the effects of correlation has been exclusively on the validity of various multiple testing procedures under dependency. For example, Benjamini and Yekutieli (2001) and Wu (2008) showed that the FDR is controlled at the nominal level by the step-up procedure (Benjamini and Hochberg, 1995) and the adaptive $p$-value procedure (Benjamini and Hochberg, 2000; Storey, 2002; Genovese and Wasserman, 2004) under different dependency assumptions. While the validity issue is important, the efficiency issue is arguably more important.

Intuitively it is clear that the dependency structure among hypotheses is highly informative in simultaneous inference and can be exploited to construct more efficient tests. For example, in comparative microarray experiments, it is found that changes in expression for genes can be the consequence of regional duplications or deletions, and significant genes tend to appear in clusters. Therefore, when deciding the significance level of a particular gene, the observations from its neighborhood should also be taken into account. It is still an open problem how to accommodate the correlation for the construction of valid and efficient multiple testing procedures.

## 4. CONCLUDING REMARKS

The two-group mixture model and the empirical Bayes approach together provide a useful general framework for multiple testing. The Lfdr, not the $p$-value, is a fundamental quantity for large-scale multiple testing. The problem of estimating the null density and the proportion of the nonnulls is important to simultaneous multiple testing. This paper raises many important questions and will definitely stimulate new research in the future. I thank Professor Efron for his clear and imaginative work.

## ACKNOWLEDGMENT

Research supported in part by NSF Grant DMS-06-04954.